\documentclass[prd,aps,tightenlines,a4paper,12pt]{revtex4}

\usepackage{graphicx}% Include figure files
\usepackage{bm}% bold math

\usepackage{enumitem}
\usepackage{url}
\usepackage{amsmath}

\newcommand{\OO}[1]{{\mathcal O}(c^{-#1})}
\newcommand{\arcsec}[0]{\hbox{$^{\prime\prime}$}}

\newcommand{\ve}[1]{\mbox{\boldmath$#1$}}

\def\pr{Phys.~Rev.}%
\def\prd{Phys.~Rev.~D}%
\def\aap{A\&A}%
\def\prl{Phys.~Rev.~Lett.}%
\def\aj{Astron.J.}%

\def\iner{{\rm iner}}

\def\KS{{\small KS}}

\begin{document}

\title{Parametrized post-Newtonian equations of motion of $N$ mass monopoles with the SEP violation}

\author{Sergei A. \surname{Klioner}} 

\affiliation{Lohrmann-Observatorium, Technische Universit\"at Dresden, 01062 Dresden, Germany}

\date{\today}

\begin{abstract}
Post-Newtonian equations of motion of a system of $N$ mass monopoles
in the framework of the PPN formalism with two parameters $\beta$ and
$\gamma$ are derived for the case when the Strong Equivalence
Principle can be violated. The derivation is based on the previously
published general framework. The multipole moments of each body are
defined as a PPN-generalization of the corresponding Blanchet-Damour
multiple moments in a relevant local reference system defined for that 
body.  The classical ten integrals of the derived equations of motion
are given. A special version of the equations of motion, for which
seven of these integrals are exact, is discussed. The derived
equations of motion can be used to test the Strong Equivalence
Principle in various solar system experiments.
\end{abstract}


\maketitle

%%% preprint
%\newpage

%%% preprint
%\tableofcontents

%%% preprint
%\newpage

\section{Introduction}

Recent advances in observational technique and future accurate stellar
catalogues will soon drastically improve the accuracy of routine
observations of asteroids and other minor bodies of the solar system.
Within a decade from now the accuracy of such observations are
expected to achieve an accuracy level of several milliarcseconds. Even
more accurate observations will be available from the ESA
second-generation astrometric mission Gaia (see e.g., \cite{ESA:2000}
and \cite{Perryman:et:al:2001}) that is expected to achieve an
accuracy between 0.2 and 3 milliarcseconds depending on the apparent
brightness of the asteroid. This gives an improvement by a factor of
100--500 compared to the typical accuracy of classical Earth-bound
positional observations of asteroids (about 1\arcsec). Moreover, Gaia
is expected to provide routinely observations of about 500000
asteroids that will allow
one to boost our knowledge of the short-term dynamics of the Solar
system.

One of the interesting applications of the drastically improved
accuracy is the use of asteroid motion to test various aspects of
relativity. Already 4 years after the discovery of asteroid Icarus,
\citet{Gylvarry:1953} has suggested to use its motion to test
general-relativistic perihelion precession. This idea has been used
several times
\citep{LieskeNull:1969,ShapiroAshSmith:1968,Shapiro:etal:1971,Sitarski:1992,Zhang:1994,ShahidSalessYeomans:1994}
and led to an independent determination of the relativistic perihelion
precession with a precision of currently 4\%. Although the perihelion
precession of Icarus (10.05\arcsec\ per century) is significantly
smaller than that for Mercury ($\sim$43\arcsec\ per century) it has
been recognized already by \citet{Dicke:1965} that asteroids with
their large inclinations and their range of semi-major axes allow one
to distinguish between the general-relativistic perihelion precession
and possible effects due to the solar oblateness (quadrupole) while it
is well known that such a distinction is virtually impossible if only
the motion of Mercury is considered. Although in recent years the
analysis of motion of the whole system of inner planets did allow to
determine separately the solar quadrupole moment and the relativistic
precession \citep{Pitjeva:2005} it remains unclear how reliable these
estimates are.

One additional aspect of the solar system dynamics is related to the
possible violation of the Strong Equivalence Principle (SEP). Solar
system dynamics was used in the very first quasi-empirical
demonstration of the so-called Nordtvedt effect which is directly
related to possible violations of the SEP \citep{Nordtvedt:1968a}.
This phenomenological approach was used in
\citep{OrellanaVucetich:1988,OrellanaVucetich:1992,PlastinoVucetich:1992}
for some tests of the SEP using asteroid motion.  A more rigorous
discussion of this phenomenon in the motion of asteroids is still to
be done. In this paper we discuss the equations of motion of $N$ mass
monopoles in the Parametrized Post-Newtonian (PPN) framework with parameters $\beta$ and
$\gamma$. After a short discussion of the Newtonian framework in Section 
\ref{section-empirical} we derive and discuss the rigorous post-Newtonian
equations of motion in Section \ref{Section-PPN-equations}.

\section{The equations of motion in the Newtonian framework}
\label{section-empirical}

We start with purely empirical Newtonian considerations. We consider
Newtonian $N$-body problem and assume that for each body $A$ we have two
different masses: the inertial mass $M_A^\iner$ appearing in the Newtonian
second law and the gravitational mass $M_A$ appearing in the formula
for the attractive force. Then the equations of motion read
\begin{equation}
\label{Nbody-Newton-eqm}
{\ddot x}_A^i=-{f_A}\,\sum_{B\ne A}\mu_B{r_{AB}^i\over r_{AB}^3}\,,
\end{equation}
\noindent
where $x_A^i$ is the position of body A, $r^i_{AB}=x^i_A-x^i_B$,
$\mu_A=GM_A$ is the mass parameter of body A, and $f_A={M_A\over
  M_A^\iner}$ is the ratio of gravitational and inertial masses of
body $A$. We see that the motion of body $A$ in an inertial reference
system depends on gravitational mass parameters $\mu_B$ of other
bodies and the mass ratio $f_A$ of body $A$.  The mass rations $f_B$
of other bodies play no role here. 

We assume in this Section that $f_A={\rm const}$ for
each body. The Lagrange function of these equations obviously read
\begin{equation}
\label{Nbody-Newton-Langrange-function}
L={1\over 2}\,\sum_Af_A^{-1}\mu_A{\dot x}_A^2-{1\over 2}\,\sum_{A}\sum_{B\neq A}{\mu_A\mu_B\over r_{AB}}\,.
\end{equation}
\noindent
From this Lagrange function it is clear that among ten classical integrals
only the integral of energy involves both gravitational and inertial masses.
The other nine integrals involve only inertial masses.
For the dynamical modeling of the $N$-body problem the mass center integral 
plays an important role since it defines the origin of a convenient inertial reference system 
to be used. The condition that the mass center of the system coincides
with the origin of the reference system involves only inertial masses
$M_A^\iner=f_A^{-1}\,M_A$ and read:
\begin{eqnarray}
\label{Nbody-Newton-center-of-mass-position}
\sum_Af_A^{-1}\,\mu_A\,x_A^i&=&0\,,
\\
\label{Nbody-Newton-center-of-mass-velocity}
\sum_Af_A^{-1}\,\mu_A\,{\dot x}_A^i&=&0\,.
\end{eqnarray}
\noindent
This condition with $f_A=1$ for all bodies is usually applied
to the solar system to define the barycentric coordinates. 
Either these conditions must be satisfied by positions and
velocities of all bodies at some initial moment of time or 
Eqs. (\ref{Nbody-Newton-center-of-mass-velocity})--(\ref{Nbody-Newton-center-of-mass-position}) can be used to eliminate one body (e.g., the Sun)
from (\ref{Nbody-Newton-eqm}).

Eq. (\ref{Nbody-Newton-eqm}) shows that if the SEP is assumed implying
$f_A=1$ for all bodies, the equations of motion of the body under
consideration are the same as in the Newtonian theory.  In the
framework of the PPN formalism, one can assume that $f_A=1$ for
laboratory test bodies and minor planets (see Section
\ref{Section-PPN-equations})
\footnote{ Let us note that in some other alternative theories of
  gravity $f_A-1$ can be relatively large even for minor solar system
  bodies (see, e.g. \citet{Overduin:2000}). This would mean that a
  test of the SEP is possible directly using Eq. (\ref{Nbody-Newton-eqm}).}.  
In this case the only effect of the 
violation of the SEP in the barycentric equations of motion of a minor body 
is the change of the definition of the center of mass as given by (\ref{Nbody-Newton-center-of-mass-position}). 

Larger effects from the possible violation of the SEP should be
expected in the motion of the Sun and major planets (bodies for which
$f_A-1$ may be expected to be maximal). The changes in the positions
of these bodies, in turn, influence also the motion of minor planets
and thus, appear indirectly in (\ref{Nbody-Newton-eqm}). In order to
make these effects explicit one should consider equations of relative
motion (e.g., heliocentric).

\section{Rigorous derivation of the EIH equations with the SEP violation}
\label{Section-PPN-equations}

Our goal now is to derive the post-Newtonian equations of motion for
$N$ mass monopoles with a possible SEP violation rigorously and
without using any empirical arguments.  \citet{Klioner:Soffel:2000}
formulated a general theory of local reference systems in the
framework of the PPN formalism with parameters $\beta$ and
$\gamma$. That theory is a generalization of the Brumberg-Kopeikin and
Damour-Soffel-Xu formalisms
\citep{kopej:88,brum:kopej:89a,brum:kopej:89b,klio:voi:93,DSXI,DSXII,DSXIII}
for the case of the special subset of the PPN formalism \citep{Will:1993}. 
The theory in \citep{Klioner:Soffel:2000}
covers a number of aspects including the definition of body's
multipole moments in its own local reference system, the
post-Newtonian tidal forces, various equations of motions, etc. It is
straightforward to use the formalism from \citep{Klioner:Soffel:2000}
to derive the relevant equations of motion for the case under study.

\subsection{The equations of motions of $N$ mass monopoles}

The barycentric equations of motion of a
system of $N$ bodies characterized by their mass monopoles and spin dipoles 
were derived in Section IX.G of
\citep{Klioner:Soffel:2000}. The derivation 
\citep{Klioner:Soffel:2000} is an immediate generalization of the
derivation of the Einstein-Infeld-Hoffmann (EIH) equations given by \citet{DSXI} for general relativity.
Below we denote the equation numbers of \citep{Klioner:Soffel:2000} as ``\KS(xx)''.  
The equations of
motion \KS(9.69) are derived rigorously using the PPN definition of
Blanchet-Damour-like multipole moments and the assumptions on the
multipole structure of gravitational field of each body given by
\KS(9.48)--\KS(9.52) and the assumption that the spin of each body 
vanishes \footnote{Section IX.G of \citet{Klioner:Soffel:2000} 
contains also the equations of motion for the bodies with non-vanishing spin dipole ${\cal S}^a$. However, 
this case will not be considered here.}:
\begin{equation}
\label{Sa=0}
{\cal S}^a=\OO2.
\end{equation}
\noindent
The notations in this paper follow those of
\citep{Klioner:Soffel:2000} (see e.g.  Section II of that work). These
assumptions represent a PPN generalization of the gravitational field
of a mass monopole.  The effect of the violation of the SEP (or the Nordtvedt effect) for assumptions
\KS(9.48)--\KS(9.52) and (\ref{Sa=0}) is given by the second term
$-R^a_iQ^a_m$ on the right-hand side of \KS(9.69), where $Q^a_m$ is
given by \KS(9.58) and \KS(9.40). Let us further simplify the equation
for the ``Nordtvedt acceleration'' $-R^a_iQ^a_m$. Substituting \KS(9.40)
into \KS(9.58) and multiplying the result with $-R^a_i$ one gets
\begin{eqnarray}
\label{Nordtvedt-acceleration}
-R^a_i\,Q_a^{\rm m}&=&
{1\over c^2}\,\eta\,{\Omega_E\over {\cal M}}\,a_E^i
\nonumber\\
&&
-{1\over c^2}\,\eta\,{{\cal N}\over{\cal M}}\,\overline{Q}_{ij}a_E^j
+{1\over 6c^2}\,(1-\gamma)\,{{\cal N}\over{\cal M}}\,{\ddot a}_E^i
\nonumber\\
&&
+{1\over 6c^2}\,(1-\gamma)\,{\dot{\cal N}\over{\cal M}}\,{\dot a}_E^i
+\OO4,
\end{eqnarray}
\noindent
where $\overline{Q}_{ij}=R^a_iR^b_jQ_{ab}$ is the tidal quadrupole
of the external gravitational field given by \KS(7.3)
and projected onto the local spatial axes. First, note that the simplified 
Nordtvedt acceleration does not depend on $\ddot{\cal N}$ while
from \KS(9.58) and \KS(9.40) one may assume such a dependence. 
Second, since ${\cal N}=\int_V\Sigma\,X^2d^3X$ (as given by \KS(8.13)\,) one has
\begin{eqnarray}
\label{dot-N-P}
\dot{\cal N}&=&2\,{\cal P},
\nonumber\\
\label{P}
{\cal P}&=&\int_V\Sigma^aX^a\,d^3X.
\end{eqnarray}
\noindent
Note that one needs these relations only in Newtonian approximation.
Therefore, one can formulate one more assumption for the structure of the 
gravitational field of
the bodies that should hold together with assumptions \KS(9.48)--\KS(9.52) and
(\ref{Sa=0}):
\begin{equation}
\label{P=0}
{\cal P}=\OO2.
\end{equation}
\noindent
With this assumption the last term in (\ref{Nordtvedt-acceleration})
vanishes. Third, ${\cal N}$ is related to Newtonian moment of inertia
of the body and ${\cal M}=\int_V\Sigma\,d^3X+\OO2$ is the body's mass
in Newtonian limit.  One can, therefore, always write ${\cal
  N}=k\,{\cal M}\,L^2$, where $L$ is the radius of a sphere
encompassing the body and $k$ is a numerical coefficient, $k\sim1$ for
small bodies and significantly smaller for giant planets and the Sun.
Using numerical characteristics of the solar system bodies one can
demonstrate that both terms in (\ref{Nordtvedt-acceleration})
proportional to ${{\cal N}\over{\cal M}}$ 
are smaller than $10^{-16}$ of the Newtonian barycentric
acceleration of the body (factors $\eta$ and $1-\gamma$ are assumed to
be of order 1 in this estimate). This is significantly smaller than the
first term in (\ref{Nordtvedt-acceleration}), which is between
$10^{-11}$ (for the Moon) and $10^{-5}$ (for the Sun) of the Newtonian
acceleration (again $\eta$ is ignored in this estimate).

In this way, we give a rigorous derivation of the EIH equations with
the single effect from the violation of the SEP -- the effect related to $f_A$ in the equations of 
motion below:
\begin{eqnarray}
\label{eom-with-eta}
\ddot{x}_A^i&=&
-f_A\,\sum_{B\ne A} \mu_B\,{r_{AB}^i\over r_{AB}^3}
\nonumber \\
&&
+{1\over c^2}\,\sum_{B\ne A} \mu_B\,
{r_{AB}^i\over r_{AB}^3}\,
\Biggl\{
(2\gamma+2\beta+1) {\mu_A\over r_{AB}}
+(2\beta-1) \sum_{C\ne A,B} {\mu_C\over r_{BC}}
+2(\gamma+\beta) \sum_{C\ne A} {\mu_C\over r_{AC}}
\nonumber \\
&&
\phantom{
+{1\over c^2}\,\sum_{B\ne A} \mu_B\,
{r_{AB}^i\over r_{AB}^3}\,
\Biggl\{ }
+{3\over 2} {{\left(r_{AB}^j \dot x_B^j\right)}^2\over r_{AB}^2}
-{1\over 2}\sum_{C\ne A,B} \mu_C\,{r_{AB}^j\,r_{BC}^j\over r_{BC}^3}
\nonumber \\
&&
\phantom{
+{1\over c^2}\,\sum_{B\ne A} \mu_B\,
{r_{AB}^i\over r_{AB}^3}\,
\Biggl\{ }
-(1+\gamma)\,\dot x_B^j\,\dot x_B^j
-\gamma\,\dot x_A^j\,\dot x_A^j
+2(1+\gamma)\,\dot x_A^j\,\dot x_B^j
\Biggr\}
\nonumber \\
&&+{1\over c^2}\,\sum_{B\ne A} \mu_B\,
{r_{AB}^j\over r_{AB}^3}\,
\biggl\{
 2(1+\gamma)\,\dot x_A^j-(2\gamma+1)\,\dot x_B^j\biggr\}\,
(\dot x_A^i-\dot x_B^i)
\nonumber \\
&&-{1\over c^2}\,\left(2\gamma+{3\over 2}\right)\,
\sum_{B\ne A}{\mu_B\over r_{AB}}\,
\sum_{C\ne A,B} \mu_C\,{r_{BC}^i \over r_{BC}^3}+\OO4\,,
\\
\label{fA-PPN}
f_A&=&1+{1\over c^2}\,\eta\,{\Omega_A\over {\cal M}_A}+\OO4\,,
\\
\label{muA}
\mu_A&=&G\,{\cal M}_A\,.
\end{eqnarray}
\noindent
These equations are valid in the PPN formalism with parameters $\gamma$ 
and $\beta$ for a system of $N$ bodies, the gravitational fields of which 
satisfy assumptions \KS(9.48)--\KS(9.52),
(\ref{Sa=0}), and (\ref{P=0}). The second and third terms 
on the right-hand side of (\ref{Nordtvedt-acceleration})
are neglected here because of their numerical smallness for applications
in solar system.

These equations of motion are in a nice agreement with empirical considerations given in Section
\ref{section-empirical}.

Note that the acceleration $\ddot{\ve{x}}_A$ of body $A$ as given by
(\ref{eom-with-eta}) depend on the mass of the body itself
$\mu_A$. Those ``self-terms'' are explicitly shown in
(\ref{eom-with-eta}) as proportional to $\mu_A$ in the first term in
the curly braces. If the motion of a minor body is considered, its gravitational 
influence of the motion of the massive bodies can be neglected and this term 
in (\ref{eom-with-eta}) can be omitted.

Eq. (\ref{fA-PPN}) gives the ratio between the gravitational and
inertial masses of body $A$ in the framework of PPN formalism, $\eta$
being the Nordtvedt parameter ($\eta=0$ if the SEP is satisfied). In the PPN formalism with two parameters
$\gamma$ and $\beta$ considered by \citet{Klioner:Soffel:2000} one has
$\eta=4\gamma-\beta-3$. A more general situation was discussion
e.g. by \citet{Will:1993}.

\subsection{The Lagrange function for the $N$-body problem}

The equations of motion (\ref{eom-with-eta}) are equivalent to the following
Lagrange function
\begin{eqnarray}
\label{lagrange-function}
&&
L = \frac{1}{2}\sum\limits_A f_A^{-1}\,\mu_A\,\dot{\ve{x}}_A^2 \left( 1 +
  \frac{1}{4c^2}\,\dot{\ve{x}}_A^2 \right) +
\frac{1}{2}\sum\limits_A {\sum\limits_{B \ne A}
  \frac{\mu_A\mu_B}{r_{AB}} 
\left( 1 + \frac{2\gamma + 1}{c^2}\dot{\ve{x}}_A^2 \right.} 
\nonumber\\ 
&&
\quad \quad \quad
\left. { 
- \frac{4\gamma + 3}{2c^2}\;\dot{\ve{x}}_A \cdot \dot{\ve{x}}_B 
- \frac{1}{2c^2}\frac{\dot{\ve{x}}_A \cdot {\ve{r}}_{AB}}{r_{AB}}
  \,\frac{\dot{\ve{x}}_B \cdot {\ve{r}}_{AB}}{r_{AB}} 
- \frac{2\beta-1}{c^2}\sum\limits_{C \ne A} \frac{\mu_C}{r_{AC}}
} \right)\;.
\end{eqnarray}
\noindent
Eq. (\ref{eom-with-eta}) can be derived from (\ref{lagrange-function}) 
up to the terms $\OO4$ using
\begin{eqnarray}
\label{Lagrange-equations}
&&
\frac{d}{dt}\frac{{\partial L}}{{\partial \dot x_A^i}} - \frac{{\partial L}}{{\partial x_A^i}} = 0\,.
\end{eqnarray}

\subsection{Integrals of motion}

Using
\begin{eqnarray}
&&
\sum\limits_A \frac{\partial L}{\partial \dot x_A^i}  = P^i = {\rm const}\,,
\\
&&
\sum\limits_A \frac{\partial L}{\partial \dot x_A^i} \,\dot x_A^i - L = h = {\rm const }\,,
\\
&&
\sum\limits_A {\varepsilon_{ijk}}\,x_A^j\frac{\partial L}{\partial \dot x_A^k}  = c^i = {\rm const}
\end{eqnarray}
\noindent
it is easy to demonstrate that the equations of motion have the following ten classical integrals
defined here with the post-Newtonian accuracy. Here we add one more assumption: the gravitation
binding energy of each body is constant
\begin{equation}
\label{Omega=const} 
\Omega_A=-{1\over 2}\,G\,\int_A\int_A {\Sigma(T,\ve{X})\,\Sigma(T,\ve{X}^\prime)
\over \left|\ve{X}-\ve{X}^\prime\right|}\,d^3X\,d^3X^\prime={\rm const}.
\end{equation}
\noindent
This means that for each body $f_A={\rm const}$. The six integrals of the center of mass read
\begin{eqnarray}
\label{center-of-mass-velocity}
&&
\sum\limits_A f_A^{-1}\,\mu_A \dot{\ve{x}}_A
\left(1 + \frac{1}{2c^2}\left(\dot{\ve{x}}_A^2 
- f_A\,\sum\limits_{B\ne A} \frac{\mu_B}{r_{AB}} \right) \right) 
\nonumber\\ 
&&
\quad\quad\quad 
- \frac{1}{2c^2}\sum\limits_A \sum\limits_{B \ne A}
  \frac{\mu_A\,\mu_B}{r_{AB}^3}\left(\ve{r}_{AB} \cdot \dot{\ve{x}}_A \right)\,\ve{r}_{AB}
= {\bf P} = {\rm const}\,,
\\
\label{center-of-mass-position}
&&
\sum\limits_A f_A^{-1}\,\mu_A \ve{x}_A\left( 1 + \frac{1}{2c^2}\left( \dot{\ve{x}}_A^2 
- f_A\,\sum\limits_{B \ne A} \frac{\mu_B}{r_{AB}}  \right) \right)  + \OO4 
= {\bf P}\,t + {\bf Q}\,,\quad {\bf Q} = {\rm const}\,.
\end{eqnarray}
The integral of energy reads
\begin{eqnarray}
\label{energy}
&&
\frac{1}{2}\sum\limits_A f_A^{-1}\,{\mu_A}\,\dot{\ve{x}}_A^2 
\left(1+\frac{3}{4c^2}\dot{\ve{x}}_A^2 \right) -
\frac{1}{2}\sum\limits_A \sum\limits_{B \ne A}
  \frac{\mu_A\mu_B}{r_{AB}} 
\left(1-\frac{2\gamma+1}{c^2}\dot{\ve{x}}_A^2\right.  
+ \frac{4\gamma+3}{2c^2}\;\dot{\ve{x}}_A \cdot \dot{\ve{x}}_B
\nonumber\\
&&
\quad \quad 
\left.  + \frac{1}{2c^2}
\frac{\dot{\ve{x}}_A \cdot {\bf r}_{AB}}{r_{AB}}
\frac{\dot{\ve{x}}_B \cdot {\bf r}_{AB}}{r_{AB}} 
- \frac{2\beta - 1}{c^2}\sum\limits_{C \ne A} \frac{\mu_C}{r_{AC}}
 \right) = h = {\rm const}.
\end{eqnarray}
Finally, the integral of angular momentum reads
\begin{eqnarray}
\label{angular-momentum}
&&
\sum\limits_A f_A^{-1}\,\mu_A\,\ve{x}_A \times \dot{\ve{x}}_A
\left( 1 + \frac{1}{2c^2} \dot{\ve{x}}_A^2 
+\frac{2\gamma+1}{c^2}\,f_A\,\sum\limits_{B \ne A}
    \frac{\mu_B}{r_{AB}}  \right)
\nonumber \\
&&
-\frac{1}{2c^2}\sum\limits_A \sum\limits_{B \ne A}
  \frac{\mu_A\mu_B}{r_{AB}} 
\left( \left( 4\gamma + 3 \right)\ve{x}_A\times \dot{\ve{x}}_B
-\frac{\ve{x}_A \times \ve{x}_B}{r_{AB}}
\frac{\dot{\ve{x}}_B \cdot \ve{r}_{AB}}{r_{AB}} \right) 
= \ve{c} = {\rm const}.
\nonumber\\
\end{eqnarray}
\noindent
Note that when all terms in
(\ref{lagrange-function})--(\ref{angular-momentum}) that are
explicitly proportional to $c^{-2}$ are omitted (while keeping $f_A$
as a symbol), one gets the exact Newtonian Lagrange function and the
corresponding integrals of the Newtonian equations with a violation of
the SEP discussed above.  In these equations
$\mu_A=G\,M_A$ corresponds to the gravitational masses of
the bodies, while $f_A^{-1}\,\mu_A=G\,M_A^{\rm iter}$ corresponds to
the inertial masses ($f_A=M_A/M_A^{\rm iner}$).

\subsection{Exact equations of motion from the Lagrange function}

The Lagrange function (\ref{lagrange-function}) gives the equations of
motion which agree with the EIH-like equations (\ref{eom-with-eta})
only approximately up to the terms $\OO4$. Here again, when computing
the equations of motion from (\ref{lagrange-function}) and
(\ref{Lagrange-equations}), the acceleration $\ddot{\ve{x}}_A$ in the
post-Newtonian terms was replaced using the Newtonian equations of
motion.  This means again that if one considers (\ref{eom-with-eta})
as exact and integrates these equations numerically, the integrals
(\ref{center-of-mass-velocity}), (\ref{energy}), and
(\ref{angular-momentum}) are not exactly constants because of the
terms $\OO4$, which are neglected in the analytical calculations.  The
equations of motion that exactly agree with the Lagrange function
(\ref{lagrange-function}) read
\begin{eqnarray}
\label{eom-exact}
\ddot{x}_A^i&=&F_A^{-1}\,\left(
H_A^i-{1\over c^2}\,f_A^{-1}\,\dot{x}^i_A\,\dot{x}^j_A\,\ddot{x}^j_A\,
+{4\gamma+3\over 2c^2}\,\sum_{B\neq A} {\mu_B\,\ddot{x}^i_B\,\over r_{AB}}
+{1\over 2c^2}\,\sum_{B\neq A} \mu_B\,{r^i_{AB}\,\over r_{AB}^3}\,\ddot{x}^j_B\,r^j_{AB}
\right)\,,
\nonumber\\
\\
\label{HA}
H_A^i&=&
-\sum_{B\ne A} \mu_B\,{r_{AB}^i\over r_{AB}^3}
\nonumber \\
&&
+{1\over c^2}\,\sum_{B\ne A} \mu_B\,
{r_{AB}^i\over r_{AB}^3}\,
\Biggl\{
(2\beta-1) {\mu_A\over r_{AB}}
+(2\beta-1) \sum_{C\ne A,B} {\mu_C\over r_{BC}}
+(2\beta-1) \sum_{C\ne A} {\mu_C\over r_{AC}}
\nonumber \\
&&
\qquad\qquad
+{3\over 2} {{\left(r_{AB}^j\,\dot x_B^j\right)}^2\over r_{AB}^2}
-(1+\gamma)\,\dot x_B^j\,\dot x_B^j
-\left(\gamma+{1\over 2}\right)\,\dot x_A^j\,\dot x_A^j
+2(1+\gamma)\,\dot x_A^j\,\dot x_B^j
\Biggr\}
\nonumber \\
&&+{1\over c^2}\,\sum_{B\ne A} \mu_B\,
{r_{AB}^j\over r_{AB}^3}\,
\biggl\{
(2\gamma+1)\,(\dot x_A^j-\dot x_B^j)\,(\dot x_A^i-\dot x_B^i)
-\dot x_A^j\,\dot x_B^i
\biggr\}\,,
\\
\label{FA}
F_A&=&f_A^{-1}\,\left(1+{1\over 2c^2} \dot{\ve{x}}_A^2\right)+{1\over c^2}\,(2\gamma+1)\,\sum_{B\neq A}{\mu_B\over r_{AB}}\,.
\end{eqnarray}
\noindent
If the accelerations in the post-Newtonian terms in (\ref{eom-exact})
are replaced using the Newtonian approximation these equations are
equivalent to (\ref{eom-with-eta}). Because of the last three terms
in (\ref{eom-exact}) Eqs. (\ref{eom-exact})--(\ref{FA}) are implicit
with respect to the accelerations $\ddot{\ve{x}}_A$ and can be solved e.g.
by iterations using the Newtonian equations for $\ddot{\ve{x}}_A$ or
the approximation $\ddot{\ve{x}}_A=F_A^{-1}\,\ve{H}_A$ as initial
approximation.

Again all the terms in $\ddot{\ve{x}}_A$ explicitly proportional to
$\mu_A$ are explicitly shown in (\ref{HA}) as the first term in the
curly braces. This term should be omitted if the motion of a minor
body is considered so that its gravitational influence on the motion
of other bodies is neglected.

\subsection{Character of the integrals of motion}

Integrals (\ref{center-of-mass-velocity}), (\ref{energy}), and
(\ref{angular-momentum}) are exact integrals of the equations of
motion with the Lagrange function (\ref{lagrange-function}). 
Therefore, these integrals remain exactly constant if the equations of
motion (\ref{eom-exact})--(\ref{FA}) are used. Integral
(\ref{center-of-mass-position}) is derived by integrating
(\ref{center-of-mass-velocity}) and using Newtonian equations of
motion in the post-Newtonian terms to replace the accelerations. It
means that when computing (\ref{center-of-mass-position}) numerically
along solutions of the equations of motion with the Lagrange function
(\ref{lagrange-function}) the quantity is not exactly linear function
of time as specified in (\ref{center-of-mass-position}), but also has
some small non-linear deviations (basically, $\ve{Q}={\rm const}+\OO4$
and those $c^{-4}$ terms depend on time in a non-linear way).

It remains unclear if one can find a form of
(\ref{center-of-mass-position}) that is satisfied exactly with either
form of the equations of motion discussed above.

\section{Conclusions}
The derived equations of motions and their integrals can be used for
tests of the Strong Equivalence Principle in the future solar system
experiments. In particular, the equations can be useful for the tests
using high-accuracy asteroid observations from the ESA space mission
Gaia as well as the SEP tests planned as a part of the BepiColombo mission
\citep{MilaniEtAl2002}.

\acknowledgments
The author thanks L\'eo Bernus for finding and correcting an error in 
Eq. (\ref{HA}).


\begin{thebibliography}{999}

\bibitem[Brumberg, Kopeikin(1989a)]{brum:kopej:89a}
    Brumberg, V.A., Kopejkin, S.M. 1989 in {\sl Reference Frames}, ed.
    J. Kovalevsky, I.I. Mueller, B.Kolaczek, Kluwer, Dordrecht, 115

\bibitem[Brumberg, Kopeikin(1989b)]{brum:kopej:89b}
    Brumberg, V.A., Kopejkin, S.M. 1989 Nuovo Cimento, {\bf 103B}, 63

\bibitem[Damour {\it et al.}(1991)]{DSXI}
Damour, T., Soffel, M., Xu, C. 1991 Phys. Rev. D {\bf 43}, 3273

\bibitem[Damour {\it et al.}(1992)]{DSXII}
Damour, T., Soffel, M., Xu, C. 1992 Phys. Rev. D {\bf 45}, 1017

\bibitem[Damour {\it et al.}(1993)]{DSXIII}
Damour, T., Soffel, M., Xu, C. 1993 Phys. Rev. D {\bf 47}, 3124

\bibitem[Dicke(1965)]{Dicke:1965}
Dicke, R.H. 1965,
%Icarus and Relativity,
\aj, 70, 395%--396

\bibitem[ESA(2000)]{ESA:2000}
ESA, 2000,
%GAIA: Composition, Formation and Evolution of the Galaxy,
%Concept and Technology Study Report, 
ESA-SCI(2000)4, Noordwijk: European Space Agency

\bibitem[Gylvarry(1953)]{Gylvarry:1953}
Gylvarry, J.J. 1953,
%Relativity Precession of the Asteroid Icarus,
\pr, 89, 1046

\bibitem[Klioner \& Soffel(2000)]{Klioner:Soffel:2000}
Klioner, S.A., \& Soffel, M.H. 2000,
%Relativistic Celestial Mechanics with PPN Parameters,
\prd, 62, ID 024019

\bibitem[Klioner, Voinov(1993)]{klio:voi:93}
    Klioner, S.A., Voinov, A.V. 1993 Phys. Rev. D {\bf 48}, 1451

\bibitem[Kopejkin(1988)]{kopej:88}
    Kopejkin, S.M. 1988 Celestial Mechanics, {\bf 44}, 87

\bibitem[Lieske \& Null(1969)]{LieskeNull:1969}
Lieske, J.H., Null, G.W., 1969,
%Icarus and the Determination of Astronomical Constants,
\aj, 74, 297%--307

\bibitem[Milani {\it et al.} (2003)]{MilaniEtAl2002}
Milani, A., Vokrouhlick\'y, D., Villani, D., Bonanno, C., Rossi, A., 2002
%Testing general relativity with the BepiColombo radio science experiment,
\prd, 66, 082001

\bibitem[Nordtvedt(1968)]{Nordtvedt:1968a}
Nordtvedt, K., Jr. 1968,
%Equivalence Principle for Massive Bodies. I. Phenomenology.
\pr, 169, 1014%--1016

\bibitem[Orellana \& Vucetich(1988)]{OrellanaVucetich:1988}
Orellana, R.B., Vucetich, H. 1988,
%The principle of equivalence and the Trojan asteroids,
\aap, 200, 248%--254

\bibitem[Orellana \& Vucetich(1992)]{OrellanaVucetich:1992}
Orellana, R.B., Vucetich, H. 1992,
%The Nordtvedt effect in the Trojan asteroids,
\aap, 273, 313%--317

\bibitem[Overduin(2000)]{Overduin:2000}
Overduin, J.M. 2000,
%Solar system tests of the equivalence principle and constrains of higher-dimensional gravity,
\prd, 62, 102001

\bibitem[Perryman {\it et~al.}(2001)]{Perryman:et:al:2001}
Perryman, M. A. C., {\it et~al.} 2001,
%GAIA: Composition, formation and evolution of the Galaxy,
\aap, 369, 339

\bibitem[Plastino \& Vucetich(1992)]{PlastinoVucetich:1992}
Plastino, A.R., Vucetich, H. 1992,
%Resonant asteroids and the equivalence principle,
\aap, 262, 321%--325

\bibitem[Pitjeva(2005)]{Pitjeva:2005}
Pitjeva, E.V. 2005,
%Relativistic Effects and Solar Oblateness from
%Radar Observations of Planets and Spacecraft,
Astronomy Letters, 31, 340%--349

\bibitem[Shahid-Saless \& Yeomans(1994)]{ShahidSalessYeomans:1994}
Shahid-Saless, B., Yeomans, D.K. 1994,
%Relativistic effects in the motion of asteroids and comets,
\aj, 107, 1885%--1889

\bibitem[Shapiro, Ash \& Smith(1968)]{ShapiroAshSmith:1968}
Shapiro, I.I., Ash, M.E., Smith, W.B. 1968,
%Icarus: Further confirmation of the relativistic perihelion precession,
\prl, 20, 1517%--1518

\bibitem[Shapiro {\it et~al.}(1971)]{Shapiro:etal:1971}
Shapiro, I.I., Smith, W.B., Ash, M.E., Herrick, S. 1971,
%General Relativity and the Orbit of Icarus,
\aj, 76, 588%--606

\bibitem[Sitarski(1992)]{Sitarski:1992}
Sitarski, G. 1992,
%On the relativistic motion of (1566) Icarus,
\aj, 104, 1226%--1229

\bibitem[Will(1993)]{Will:1993}
Will, C. M. 1993,
Theory and experiment in gravitational physics, Cambridge: Cambridge University Press

\bibitem[Zhang(1994)]{Zhang:1994}
Zhang, J. 1994,
%A study of general relativistic effects in the motion of Icarus,
Chinese Astronomy and Astrophysics, 18, 108%--105

\end{thebibliography}
\end{document}